\documentclass[12pt]{iopart}

\usepackage{ifthen}
\usepackage{ifpdf}
\usepackage{color}

\ifpdf
\usepackage{graphicx}
\usepackage{epstopdf}
\else
\usepackage{graphicx}
\usepackage{epsfig}
\fi
\graphicspath{{./Figs/}{./}}

\usepackage{latexsym}
\usepackage{amssymb}
\usepackage{bm}

\newcommand{\tbox}[1]{\mbox{\tiny #1}}

\newcommand{\eexp}[1]{\mathrm{e}^{#1}}

\newcommand{\ket}[1]{\left| #1 \right\rangle}
\newcommand{\braket}[2]{ \left\langle #1 \middle| #2 \right\rangle}
\newcommand{\Braket}[3]{ \left\langle #1 \middle| #2 \middle| #3 \right\rangle}

\newcommand{\be}[1]{\begin{eqnarray}\ifthenelse{#1=-1}{\nonumber}{\ifthenelse{#1=0}{}{\label{e#1}}}}
\newcommand{\beq}{\begin{eqnarray}}
\newcommand{\eeq}{\end{eqnarray}} 

\newcommand{\hide}[1]{}
\newcommand{\Eq}[1]  { {\textcolor{blue}{Eq.}}~(\ref{#1}) } 
\newcommand{\Fig}[1] { {\textcolor{blue}{Fig.}}~\ref{#1} }
\newcommand{\sect}[1]{{\bf #1.-- }}

\newcommand{\mycite}[1]{\textcolor{blue}{\cite{#1}}}

\newcommand*{\ddt}{\frac{\texttt{d}}{\texttt{d}t}}
\newcommand*{\ppt}{\frac{\partial}{\partial t}}

\newcommand*{\ha}{\hat{a}}

\newcommand*{\hn}{\hat{n}}

\begin{document}

\title[Quantum thermalization]
{Quantum thermalization: anomalous slow relaxation due to percolation-like dynamics}

\author{Christine Khripkov$^{1}$, Amichay Vardi$^{1}$, Doron Cohen$^{2}$}

\address{
\mbox{$^1$Department of Chemistry, Ben-Gurion University, Beer-Sheva 84105, Israel} 
\mbox{$^2$Department of Physics, Ben-Gurion University, Beer-Sheva 84105, Israel}}

\begin{abstract}
We highlight a dynamical anomaly in which the rate of relaxation towards thermal equilibrium in a bi-partite quantum system violates the standard linear-response (Kubo) formulation,  even when the underlying dynamics is highly chaotic. This anomaly originates from an $\hbar$-dependent sparsity of the underlying quantum network of transitions. Using a minimal bi-partite Bose-Hubbard model as an example, we find that the relaxation rate acquires an anomalous $\hbar$ dependence that reflects percolation-like dynamics in energy space. 
\end{abstract}


The connection between thermalization and chaotic ergodicity is well-established for classical systems~\mycite{Dorfman}. 
Since strict dynamical chaos is absent in isolated quantum systems, contemporary research efforts 
\mycite{Kinoshita06,Trotzky12,Gring12,Deutsch91,Srednicki94,Yukalov11,Polkovnikov11,Rigol08,Cassidy09,Eckstein08,Pal10,Ponomarev11,Bunin11,Ates12,Tikhonenkov13,Niemayer13,Niemayer14,Bartsch08}
are aimed to find novel quantum signatures such as Anderson localization \mycite{Fleishman80,Basko06,BarLev14} in the thermalization of quantized chaotic systems.
The current paradigm for thermalization of coupled quantum subsystems is Linear Response Theory (LRT). 
If the underlying classical dynamics is chaotic, thermalization is attained via diffusive spreading 
which is described by a Fokker-Planck-Equation (FPE) \mycite{Bartsch08,Bunin11,Ates12,Tikhonenkov13,Niemayer13,Niemayer14}, 
leading to ergodization of the composite system over all accessible states within a microcanonical energy shell. 

LRT is related to the Fermi-golden-rule (FGR) picture  
in which the rates of transitions between the unperturbed eigenstates 
of the subsystems are given by first-order-perturbation matrix elements, 
but over long timescales that involve many perturbative orders. 
The diffusion coefficient~$D$ of the FPE is estimated from these rates 
by a Kubo formula \mycite{Wilk88,Wilk95}.
LRT implies quantum-to-classical correspondence (QCC) in the FPE description,  
which is somewhat analogous to the Thomas-Reiche-Kuhn $f$-sum-rule, 
and has been termed `restricted QCC' \mycite{Cohen99}. 
The argument that supports restricted QCC with regard to the FPE picture 
is based on the observation that for short times the {\em variance} (unlike the higher moments) 
features a robust QCC, while for long times the central limit theorem makes all higher moments irrelevant. 
Thus LRT based description becomes accurate far beyond the naive expectation. 
The restricted QCC assumption prevails in all current work on thermalization \mycite{Deutsch91,Srednicki94,Yukalov11,Polkovnikov11,Rigol08,Cassidy09,Eckstein08,Pal10,Ponomarev11,Bunin11,Ates12,Tikhonenkov13,Niemayer13,Niemayer14}.

Deviations from LRT have either a classical or a quantum origin. Classical deviations result from dynamical quasi-integrability in the mixed phase space \mycite{Yurovsky11,Santos12} which can make thermalization a slow and intricate process \mycite{Kinoshita06,Trotzky12,Gring12}. By contrast {\em quantum anomalies} are directly related to the breakdown of QCC due to the finite value of the Planck constant~$\hbar$. One well-known example for such quantum anomaly is the loss of ergodicity due to many-body Anderson localization~\mycite{Fleishman80,Basko06,BarLev14}.

In this Letter we highlight a new type of quantum anomaly which does not originate from the lack of quantum ergodicity, but from the $\hbar$-dependent sparsity of the quantum network of transitions. The classical Kubo-FGR picture relies critically on the existence of a dense, connected network of transitions between all the available states, so that all transitions contribute to the diffusive energy spreading process. However, such dense networks do not always exist. The quantum network of transitions is generally {\em sparse} \mycite{slk}, resulting in a {\em percolation}-like process of energy spreading, that is dominated by bottlenecks and preferred pathways. As a result, the Kubo formula grossly overestimates the thermalization rate and QCC is lost even when the underlying classical dynamics is highly chaotic.

To illustrate this point, we consider a minimal Bose-Hubbard model of a bi-partite $N$-boson system,  where $\hbar=1/N$ plays the role of the Planck constant. We show that while the thermalization process is still described by the FGR picture, resulting in an FPE, it involves an anomalous $\hbar$-dependent diffusion-coefficient~$D$ whose estimate requires a resistor-network calculation. Thus, while the approach to equilibrium still relies on diffusive energy flow with the same long-time stationary energy distributions, the unique mechanism of 'quantum thermalization via percolation' can be much slower than its classical counterpart.
Further (technical) details regarding the resistor network calculation; the {\em percolation}-like aspect; and its $\hbar$ dependence, are provided in the appendices.

\section{Model system}

\begin{figure}

\centering
\setlength{\unitlength}{1mm}
\begin{picture}(90,76)
\put(0,0){\includegraphics[width=88mm]{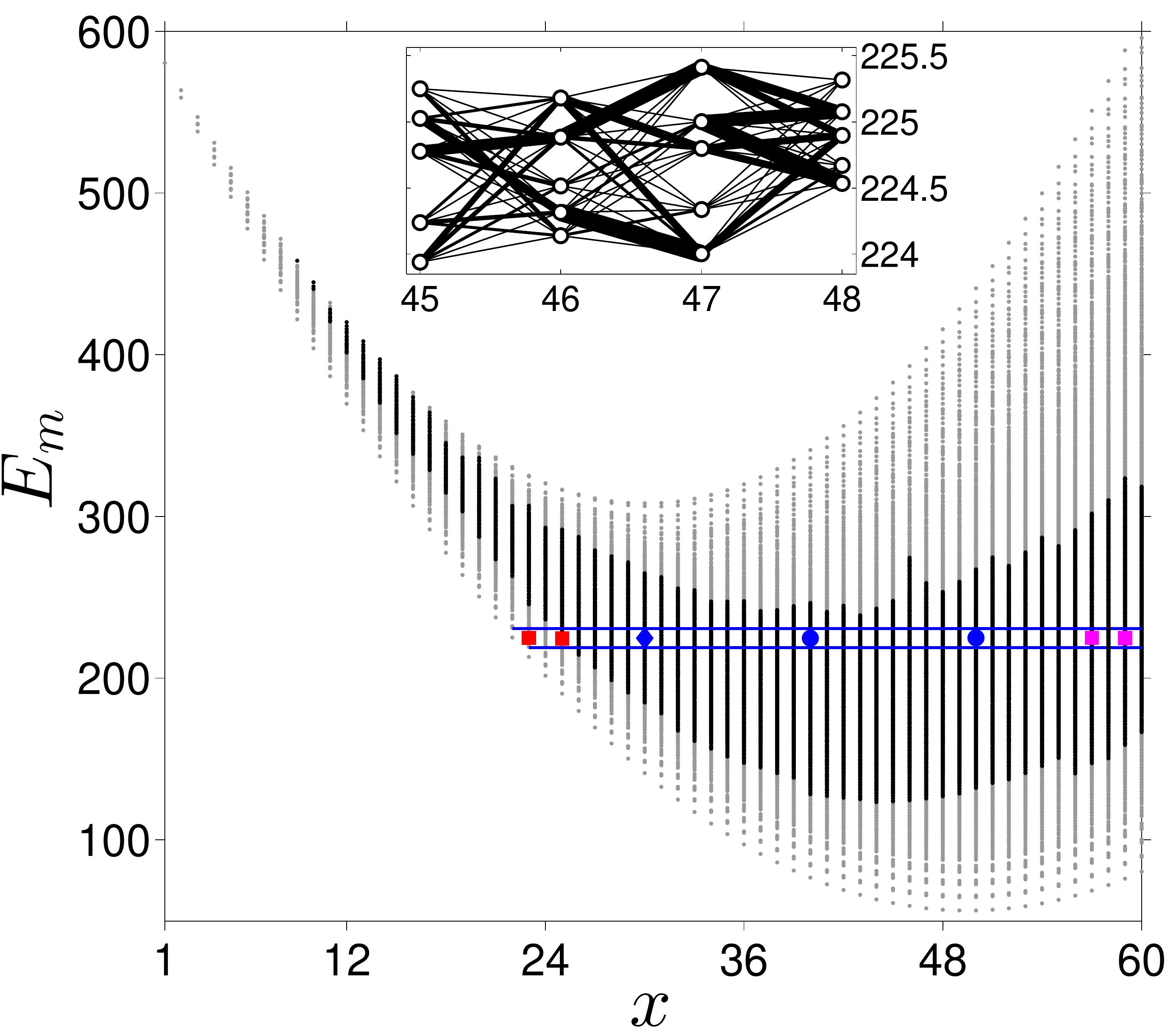}}
\put(14,10){\includegraphics[width=24mm]{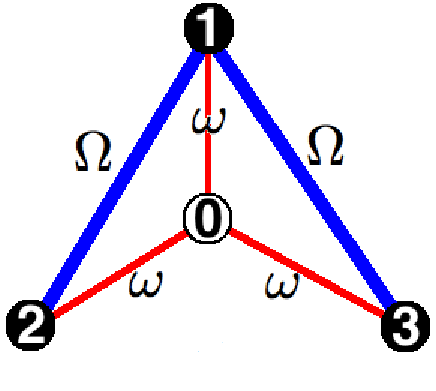}}
\end{picture}

\caption{\label{fig:energyshell}
{\bf Quantum network of transitions.} 
The trimer-monomer model system is schematically illustrated in the lower left inset.
In the absence of trimer-monomer coupling the energy eigenstates  
can be classified by the trimer population~$x$.
The parameters are $N=60$, $NU=20$, and $\Omega=3.17$.
The dark points mark eigenstates lying in chaotic phase-space regions. 
The blue band marks the accessible states within the 
energy window ${E_m \pm 1/\tau}$,  
where~$|m\rangle$ is the central state at the ${x=30}$ band,  
and $\tau$ is obtained from \Eq{eq:tau} with ${\omega=0.1\Omega}$. 
The diamond marker denotes the chaotic preparation for 
the simulation of \Fig{fig:diffusivedynamics}, 
whereas other markers denote the additional preparations 
used in \Fig{fig:saturations}.
The upper inset zooms over a segment of the energy shell, 
and illustrates the network of transitions 
formed by the perturbation. 
The width of each connecting line is proportional 
to the strength of the coupling matrix element.
}
\end{figure}

Consider an isolated system of $N$~bosons in four second quantized modes.
The operators $\ha_j$, $\ha_j^\dag$ and $\hn_j=\ha_j^\dag\ha_j$  
annihilate, create and count particles in site $j$. 
The dynamics is generated by the Bose-Hubbard Hamiltonian (BHH) 
\beq\label{eq:hamiltonian}
\mathcal{H} \ = \
\frac{U}{2}\sum_{j=0}^{3} \hn_j^2 - \frac{\Omega}{2} (\ha_1^\dag\ha_2 +\ha_1^\dag\ha_3 + \mbox{h.c.}) +  \mathcal{H}_P~,
\eeq
where $U$ is the on-site interaction, 
and $\Omega$ couples a chain of 
three sites ${j=1,2,3}$. 
The perturbation $\mathcal{H}_P$ generates transitions 
to an additional $j=0$ site, namely,   
\beq
\mathcal{H}_P \ = \  - \frac{\omega}{2} \sum_{j=1}^3 (\ha_0^\dag \ha_j + \mbox{h.c.})~. 
\eeq
Thus $\mathcal{H}$ describes a bi-partite system: a BHH {\em trimer} coupled to a {\em monomer} 
(see schematic illustration in \Fig{fig:energyshell}). 
Weak coupling between the two subsystems is assumed (${\omega \ll \Omega, NU}$),  
and the interaction within the trimer is quantified 
by the dimensionless interaction parameter ${u=NU/\Omega}$. 
In the classical description each site is described by conjugate action angle 
variables ${(n_j,\varphi_j)}$. The standard procedure \mycite{Chuchem10} is to work with dimensionless 
variables. In particular the scaled occupations are $n_j/N$, hence upon quantization 
the scaled Planck constant is $\hbar=1/N$. The classical limit is attained 
by taking the limit ${N \rightarrow\infty}$ keeping $NU$ constant. 
In this limit quantum fluctuations diminish and the bosonic operators 
can be replaced by c-numbers. The semiclassical description becomes valid if $\hbar\ll 1$.

The above {\em trimer plus monomer} model is the minimal
Bose-Hubbard configuration which allows chaos and thermalization, 
because the trimer subsystem is classically chaotic \mycite{Kottos}, while a dimer is not. 
Furthermore, this type of minimal configuration serves as the {\em building-block}
for progressive thermalization of large arrays \mycite{Basko,Henn}.

\section{Quantum network of transitions}

The trimer population ${\hat{x} \equiv \hn_1+\hn_2+\hn_3}$ 
commutes with the unperturbed ($\omega=0$) Hamiltonian $\mathcal{H}_0$, 
and therefore  constitutes a good quantum number in the absence of coupling.  The unperturbed spectrum as defined by the eigenstate equation $\mathcal{H}_0\ket{m}=E_m\ket{m}$ is plotted in \Fig{fig:energyshell}.  Each unperturbed eigenstate is associated with a 'position' $x_m$ on the trimer occupation grid. Thus, \Fig{fig:energyshell} should be interpreted as specifying the unperturbed trimer spectrum for all possible trimer occupations from $x=1$ to $x=N$. 
We identify the region of chaotic dynamics by a Brody parameter map \mycite{Brody81} (see Appendix~B), 
verified by classical Poincare sections (not shown). 
Eigenstates  supported by chaotic phase-space regions are marked in black in \Fig{fig:energyshell}. 

The perturbation due to coupling with the additional mode allows transfer of particles and energy and thus generates transitions along the occupation axis~$x$.  The transition strengths are given as $\Braket{n}{\mathcal{H}_P}{m}$. The upper inset of \Fig{fig:energyshell} depicts the coupling network within a narrow $[x,E]$ window. Due to the wide distribution of transition strengths, the obtained network is glassy. This glassiness is reminiscent of the sparsity that arises in integrable systems due to selection rules \mycite{slk}.

\begin{figure}
\centering
\includegraphics[width=90mm]{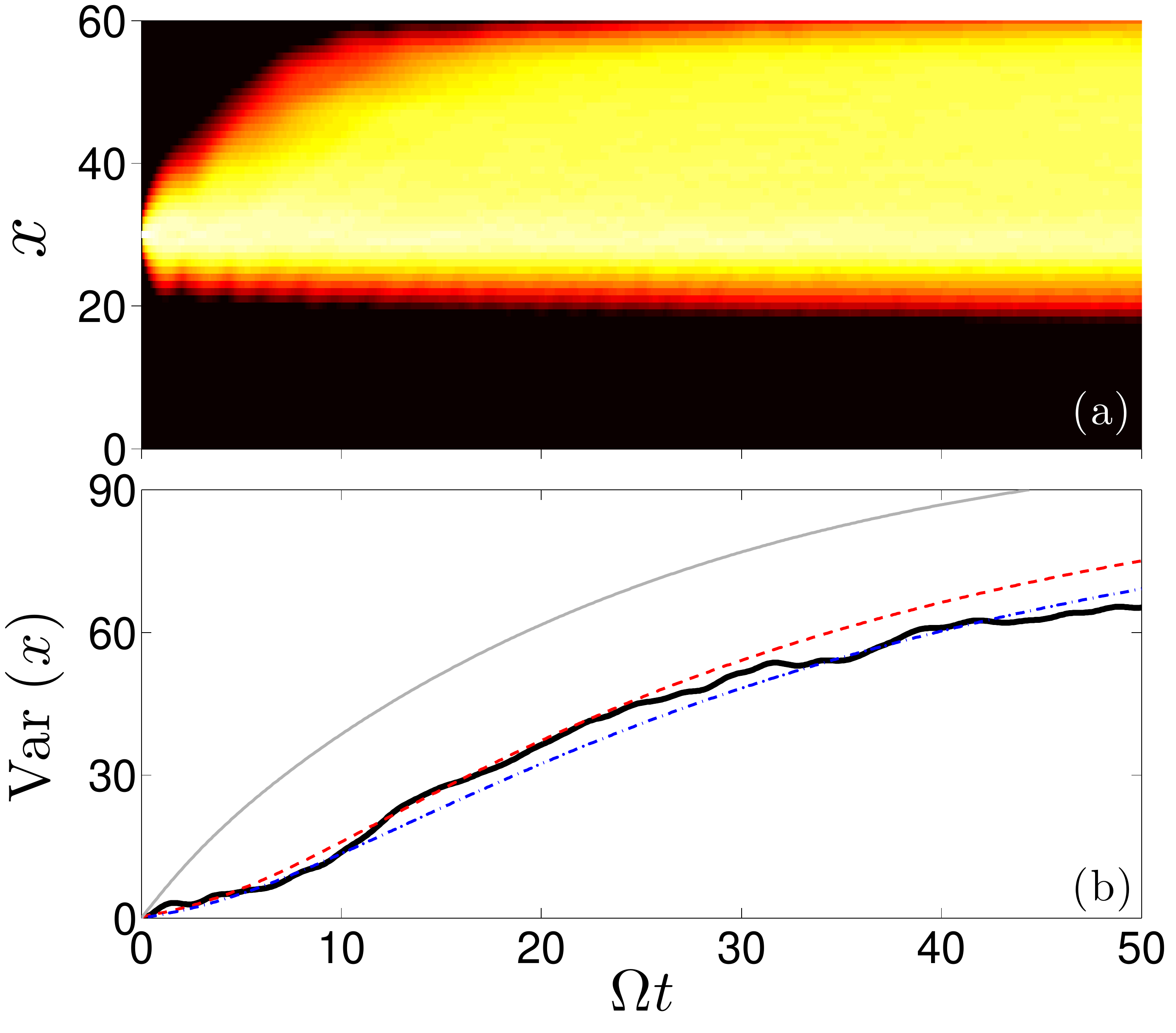}
\caption{\label{fig:diffusivedynamics}
{\bf Diffusive quantum thermalization.}
The distribution $P_t(x)$ is imaged as a function of time~(a),   
and the corresponding growth of variance is plotted using the same time axis~(b). 
In the latter the variance of the distribution (thick black line) 
is compared with the stochastic approximations.
The FGR simulation (dashed red) and the corresponding 
FPE simulation with a resistor-network estimate $D_{\tbox{qm}}(x)$ for the diffusion 
coefficient (dot-dashed blue) agree with the quantum simulation, 
unlike the traditional FPE simulation (thin solid gray) 
with a Kubo-type estimate $D_{\tbox{cl}}(x)$ for the diffusion. 
Parameters are the same as in \Fig{fig:energyshell}. 
}
\end{figure}

\section{Diffusive spreading}

We focus our attention on the evolution of the probability distribution $P_t(x)$, 
starting with an initial state $|m\rangle$. This preparation is an eigenstate of the unperturbed Hamiltonian, 
but a far from equilibrium initial state for the combined system. 
The system's parameters are chosen such that the energy of this state 
(diamond blue marker in \Fig{fig:energyshell}) lies within a broad chaotic phase-space window. 

A representative example for the evolution of the $x$ probability distribution in the chaotic regime is plotted in \Fig{fig:diffusivedynamics} with the growth of variance $\mbox{Var}(x)$ depicted in the lower panel. Similarly to the results of Refs.~\mycite{Ates12,Tikhonenkov13},  the hallmark of chaos is 
stochastic-like spreading. This diffusive behavior persists until the distribution saturates the accessible energy window, thus leading to thermalization. 

However, the rate in which the equilibrium distribution is approached is very far from the conventional Kubo estimate and is therefore highly non-classical. The thin solid gray line in the lower panel of \Fig{fig:diffusivedynamics} corresponds to the traditional FPE description of the dynamics, with a diffusion coefficient $D_{\tbox{cl}}(x)$ that corresponds to the classical result.  It is evident that the standard classical prediction greatly overestimates the equilibration rate and that indeed quantum thermalization is slower due to the sparsity of the transition network. By contrast,  the dot-dashed blue line also depicts an FPE description, but with a percolation-theory resistor network estimate $D_{\tbox{qm}}(x)$ for the diffusion coefficient, that, as described below, takes into account the $\hbar$ dependent transition network sparsity.  We thus observe a novel anomalous process of {\em quantum thermalization}, which is stochastic and adheres to an FPE description, albeit with an underlying  percolation-like spreading process which does not correspond to the classical dynamics.

\begin{figure}
\centering
\includegraphics[width=90mm]{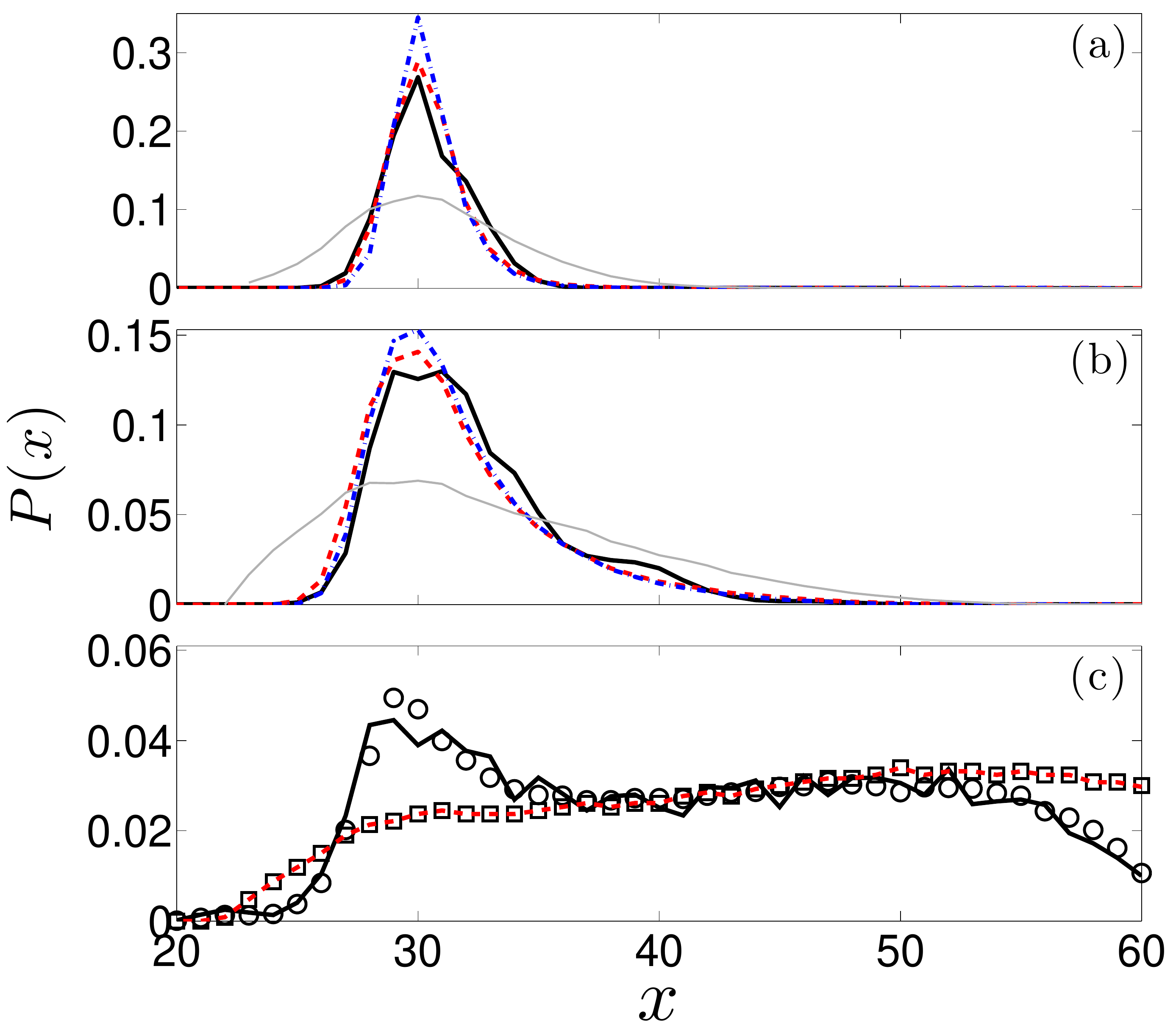}

\caption{\label{fig:snapshots}
{\bf Snapshots of the spreading profile.}
The energy probability distribution $P_t(x)$ of \Fig{fig:diffusivedynamics} is plotted at: 
(a)~$\Omega t=2.5$, 
(b)~$\Omega t=10$, 
(c)~$\Omega t=1000$.
Line types are as in \Fig{fig:diffusivedynamics}b with dashed line corresponding to FGR, 
dash-dotted line depicting the FPE propagation with $D_{qm}(x)$, and gray solid line 
depicting the FPE propagation with $D_{cl}(x)$.
Circles in panel (c) mark the saturation profile calculated 
using the convolution \Eq{eq:infty}, while squares 
mark the ergodic micro-canonical profile ${\propto \tilde g(x)}$. 
}
\end{figure}

\section{Evolution of the distribution profile}

Several snapshots of $P_t(x)$ during the thermalization process are plotted in \Fig{fig:snapshots}, showing good agreement between the percolation-FPE and the full numerical simulation of the four-mode dynamics. By contrast, the conventional classical FPE thermalization gives far broader distributions  at the same times. 

\begin{figure}[b!]
\centering
\includegraphics[width=90mm]{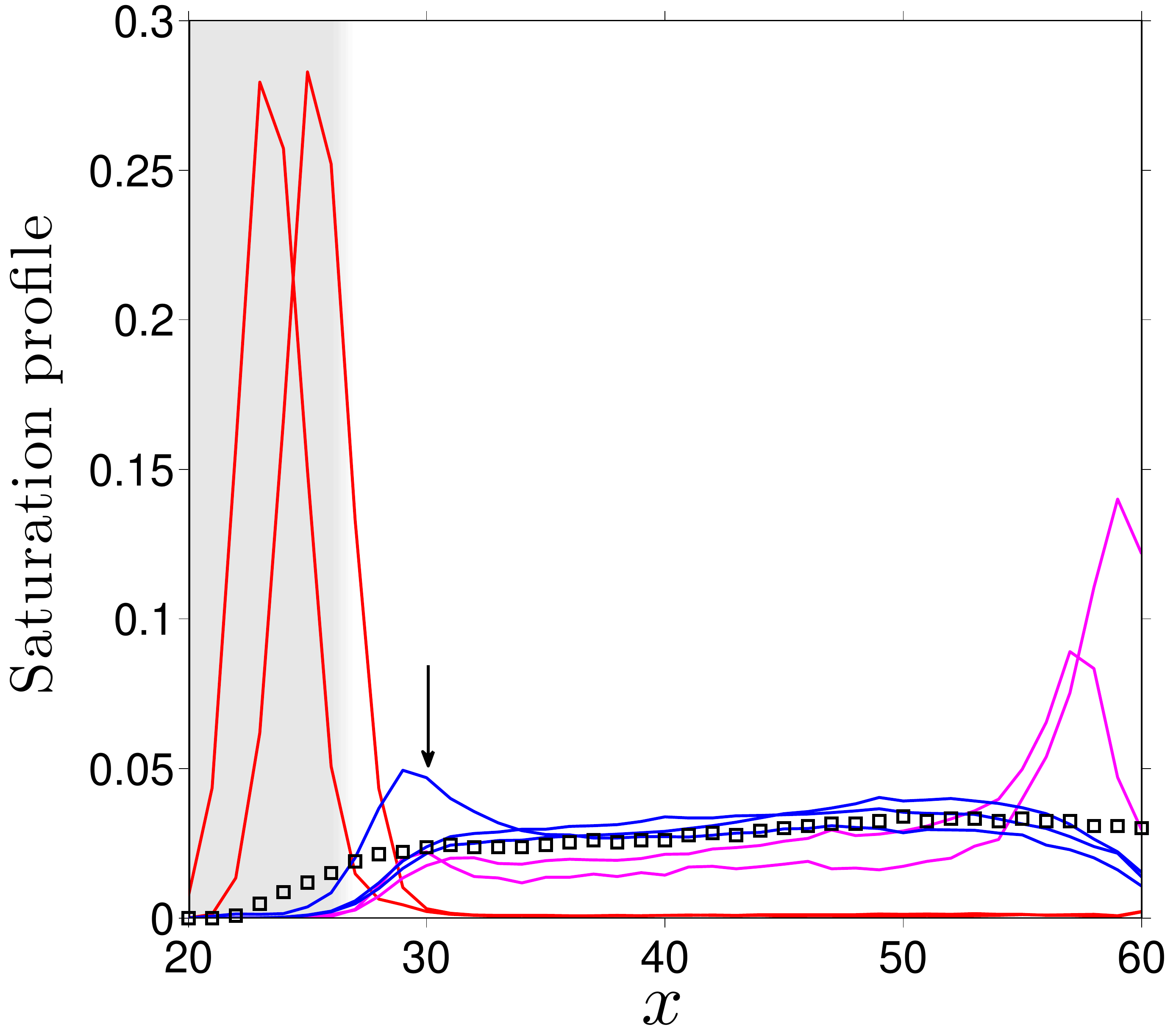}

\caption{\label{fig:saturations}
{\bf Saturation profiles.} 
Quantum saturation profiles starting from the initial states marked in \Fig{fig:energyshell}, compared to the micro-canonical ($\propto \tilde g(x)$) thermal distribution (square markers). The quasi-integrable region is marked in gray and an arrow in the chaotic region marks the initial state used in \Fig{fig:diffusivedynamics}--\Fig{fig:Dx}. Non-ergodicity is due to quasi-integrability at the low $x$ region (red lines) and due to Anderson-type localization at the high $x$ region (magenta lines). Quantum thermalization is obtained for intermediate $x$ preparations, regardless of the precise initial conditions  (blue lines).}
\end{figure}

An additional observation concerns the long time equilibrium distributions, plotted in \Fig{fig:snapshots}c. The saturation profile $P_{\infty}(x)$ of the FPE is proportional, as expected, to the density of states ${\tilde g}(x)$. By contrast the exact equilibrium distribution is somewhat non-ergodic. The lack of ergodicity in the low $x$ region of the saturation profile, is due to 
residual integrability within islands of the underlying mixed phase-space. It therefore disappears when the simulation is started deeper within the chaotic sea, see \Fig{fig:saturations}.
In addition, there are deviations from ergodicity in the high $x$ region
due to Anderson-type localization. The former semiclassical effect and the latter 
quantum anomaly are both distinct from the dynamical anomaly which constitutes our 
main theme. For further detail on these deviations see Section~\ref{sSat}.

\section{Stochastic FGR rate equations}

The transition rates between two chaotic sub-systems are non-zero 
provided ${|E_{n}-E_{m}|<1/\tau}$, where the bandwidth $1/\tau$ is 
determined by the width of the power-spectrum of the perturbation \mycite{Tikhonenkov13}.
The FGR estimate for the non-zero rates is accordingly,  
\beq\label{eq:FGRrates}
\Gamma_{mn} = 2\pi \tau \, \left|\Braket{n}{\mathcal{H}_p}{m}\right|^2~. 
\eeq
With these rates, the master equation for the occupation probabilities is 
\beq\label{eq:FGR}
\ddt p_n \ \ = \ \ - \sum_m \Gamma_{mn}(p_n - p_m).
\eeq
Our model is sub-minimal in the sense that the monomer is not 
a chaotic sub-system. Still, the dynamics is the same as for two chaotic sub-systems with 
$1/\tau$ determined by the width of the {\em energy shell}. Namely, 
\beq\label{eq:tau}
\frac{1}{\tau} \ = \ \sqrt{\Braket{m}{\mathcal{H}^2}{m}-\Braket{m}{\mathcal{H}}{m}^2}~.
\eeq
Only states within this energy shell, marked by blue lines in  \Fig{fig:energyshell}, contribute to the thermalization process. States outside it do not participate in the dynamics. The red dashed lines in \Fig{fig:diffusivedynamics} and \Fig{fig:snapshots} correspond to the propagation of \Eq{eq:FGR} (see appendix). The agreement with the full quantum simulation validates the stochastic FGR picture.

\section{The FPE description}

\begin{figure}
\centering
\includegraphics[width=90mm]{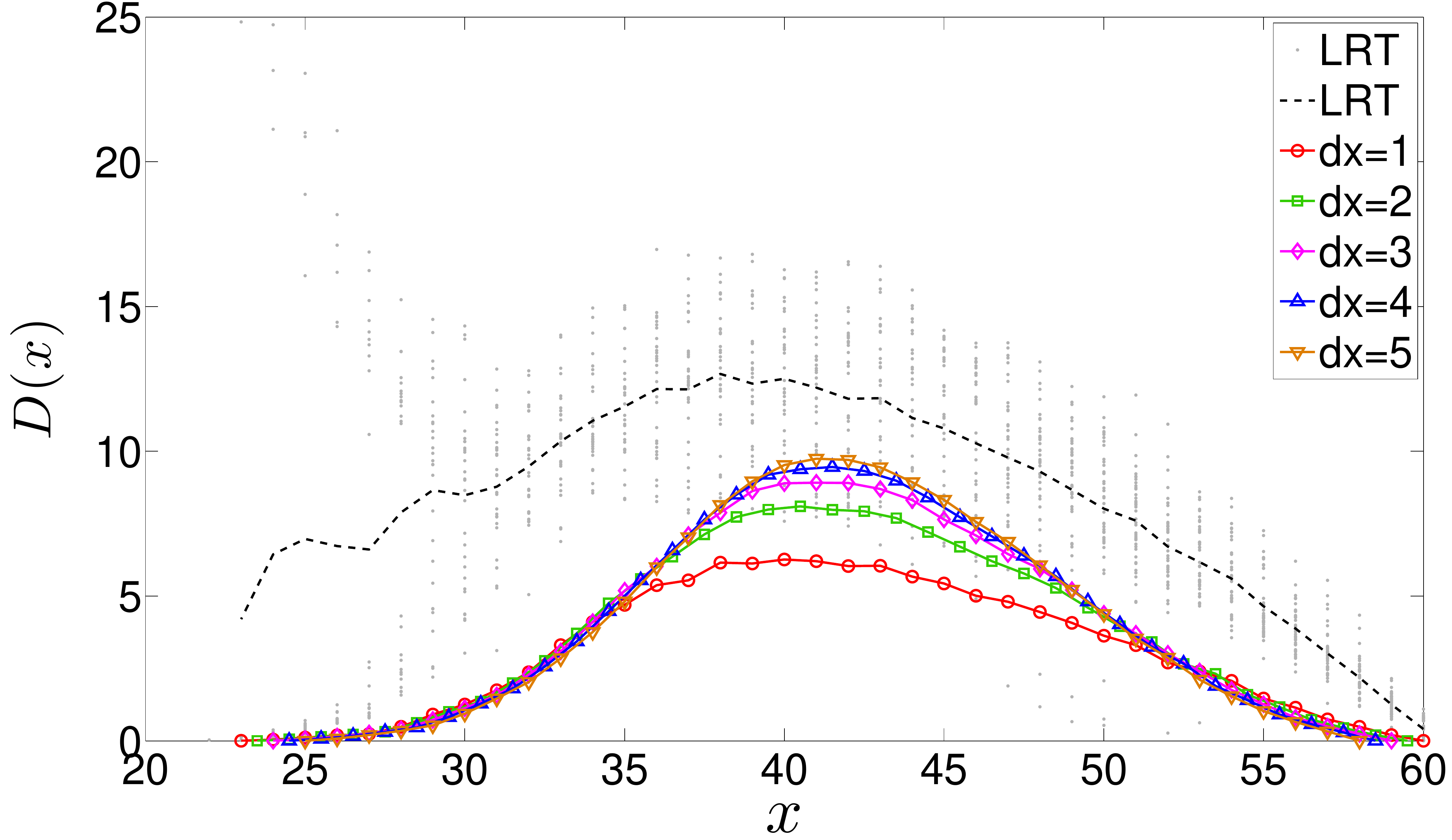}
\caption{\label{fig:Dx}
The resistor-network estimated $D_{\tbox{qm}}(x)$ 
is calculated over $dx$ segments (see Appendix~D, note convergence). 
It is contrasted with $D_{\tbox{cl}}(x)$ of the Kubo calculation: 
gray dots for each $m$ in \Eq{eq:DLRT};  
and dashed black line for the $m$-averaged result.} 
\end{figure}

Coarse graining  of the kinetic equations (\ref{eq:FGR})  
results in the FPE, which is merely a diffusion equation in $x$~space 
\beq\label{eq:FPEx}
\ppt P(x)=\frac{\partial}{\partial x}\Big[\tilde g(x) D(x) 
\frac{\partial}{\partial x} \Big(\tilde g(x)^{-1} P(x)  \Big) \Big]~.
\eeq
Here $\tilde g(x)$ is the density of states 
within the allowed energy shell. 
Unlike the textbook version of the diffusion equation,  
which assumes uniform $\tilde g(x)$ and $D(x)$,  
the form of the FPE (\ref{eq:FPEx}) 
reflects the simple observation that an ergodic distribution 
occupies uniformly all accessible eigenstates, 
so that the FPE ergodic saturation profile must satisfy $P_{\infty}(x) \propto \tilde g(x)$.
The standard linear response estimate for the diffusion 
coefficient, i.e. the Kubo formula \mycite{Wilk88,Wilk95}, 
is based on a second moment calculation: 
\beq\label{eq:DLRT}
D_{\tbox{cl}}(x) \ = \ \left\langle\ \frac{1}{2} \sum_n  (x_n-x_m)^2 \ \Gamma_{nm}  \right\rangle,
\eeq
where the brackets correspond to averaging over 
all the in-band states $m$ in the vicinity of~$x$.
The result of the $D_{\tbox{cl}}(x)$ calculation is illustrated in \Fig{fig:Dx}. 
We have verified that the obtained values of  $D_{\tbox{cl}}(x)$ are robust, 
i.e. are not sensitive to the exact value of the micro-canonical width~$1/\tau$.

\section{Resistor-network calculation}

As mentioned above, the FPE simulation with the standard diffusion coefficient $D_{\tbox{cl}}(x)$ fails to reproduce the true dynamics as illustrated in \Fig{fig:diffusivedynamics}. This striking breakdown of QCC is due to the percolation-like nature of energy spreading.  As appropriate for a percolation process, $D(x)$ should be estimated from the conductivity of the 'resistor network' that is formed by the quantum transitions \mycite{slk}. Such evaluation gives the proper weight to low-resistance, well-connected links, as opposed to the over-estimated democratic weighing of \Eq{eq:DLRT}. 
Thus, in steady state \Eq{eq:FGR} is formally the same as  Kirchhoff's equation
\beq\label{eq:kirchhoff}
\sum_m G_{mn} \, (V_n-V_m) \ \ = \ \ I_n
\eeq
where the conductances $G_{mn}$, and the voltages $V_n$,
are analogous to $\Gamma_{nm}$ and $p_n$ respectively.
In order to calculate the conductance of a small $x$~segment ${[x_1,x_2]}$, 
we set $I_n=0$ for all internal nodes, and $I_n=\pm I_{\tbox{source}}$ at the endpoints. 
The detailed numerical procedure is provided in Appendix~C. 
Solving for the voltage we deduce that the conductance 
of the $x$~segment is $G(x)= I_{\tbox{source}}/(V_2-V_1)$, 
and hence the conductivity is ${D_{\tbox{qm}}(x)=(x_2{-}x_1)G(x)}$. 

As shown in \Fig{fig:Dx}, the resistor-network calculated diffusion coefficient $D_{\tbox{qm}}(x)$ is substantially smaller than the Kubo result $D_{\tbox{cl}}(x)$.  As previously stated, the FPE simulation (Appendix~D) with $D_{\tbox{qm}}(x)$, presented in \Fig{fig:diffusivedynamics}, agrees well with the quantum simulation.   The agreement persists as long as the spreading is within the chaotic region of the energy shell, confirming our expectations.

\section{Saturation profile}
\label{sSat}

For completeness we further discuss the saturation profiles of \Fig{fig:saturations}.
Given an initial state ($m$), we take its overlap with the exact eigenstates ($\nu$), 
\beq\label{eq:LDOS}
P(\nu|m) \ \ = \ \ \left|\braket{\nu}{m}\right|^2.
\eeq
Evolving the initial state $m$ in time we define the probability distribution 
\beq\label{eq:distribution}
P_t(n|m) \ \ = \ \ \left|\Braket{n}{\eexp{-i\mathcal{H}t}}{m}\right|^2.
\eeq
The $P_t(x)$ distribution is related to this kernel by binning together the probabilities of all the unperturbed eigenstates with the same trimer occupation, namely ${P_t(x) = \sum_{n}^{(x)} P_t(n|m)}$ where the summation is over all unperturbed states $n$ with  ${x_n=x}$. Note that while $P(\nu|m)$ is the fixed probability distribution between the exact eigenstates of the composite four-mode system, $P_t(n|m)$ is the time-dependent probability distribution between the eigenstates of an uncoupled trimer-monomer subsystem.

The long time saturation profile of the evolving distribution $P_t(n|m)$, 
can be obtained directly from the overlaps $P(\nu|m)$, via the convolution formula
\beq\label{eq:infty}
P_{\infty}(n|m) \ \ = \ \ \sum_{\nu} P(\nu|n)P(\nu|m).
\eeq
This relation is obtained by expanding the states $|n\rangle$ and $|m\rangle$ of \Eq{eq:distribution} in the $|\nu\rangle$ basis, assuming that the spectrum is non-degenerate; hence only diagonal terms survive after the long time averaging \mycite{Short12,Reimann12}. Note that whenever the Wigner surmise applies, degeneracies have measure zero due to level repulsion. 
We have verified that \Eq{eq:infty} is in very good agreement with the exact simulation, as demonstrated in \Fig{fig:snapshots}c.

It thus becomes clear that the deviation from ergodicity is related 
to the localization of some unperturbed-eigenstate preparations $|m\rangle$, 
as reflected in the overlaps $P(\nu|m)$. 
Several preparations with the same energy but lying in different phase-space 
regions are marked in \Fig{fig:energyshell}, 
while their associated saturation profiles are shown in \Fig{fig:saturations}.
Preparations in the chaotic region give the micro-canonical ergodic saturation profile ${ P_{\infty}(x) \propto \tilde{g}(x) }$, 
independently of the choice of initial state (blue lines). 
In the low $x$ region of the saturation profile the localization  
is of semi-classical nature, due to the underlying mixed phase-space 
which contains remnant quasi-integrable regions. 
Preparations supported by such integrable islands have narrow $P(\nu|m)$ 
which leads to localized saturation profiles. 
At the high $x$ region, the coupling between eigenstates in different $x$ manifolds, as quantified by the value of the diffusion coefficient $D_{\rm qm}$,  becomes small (see \Fig{fig:Dx}). Consequently, the Anderson localization length $\xi=2\pi \tilde{g} D_{\rm qm}$ is only a few sites, again resulting in localized saturation profiles (magenta lines). The deviation of the saturation profile in this region from the ergodic result of the stochastic FGR calculation (see e.g. \Fig{fig:snapshots}c) indicates that this is an Anderson-type interference effect.

\section{Experimental realization}
\label{ExpReal}

Few-mode Bose-Hubbard systems can be realized in confining potentials with toroidal shapes and tunable weak links \mycite{Amico05,Henderson09,Wright13,Amico14}. Of particular relevance for the realization of bi-partite Bose-Hubbard models is the experimental generation of  arbitrary and dynamical potentials in a $^{87}$Rb Bose-Einstein Condensate by means of a rapidly moving laser beam \mycite{Henderson09}. Alternatively, the interference of the rotationally-symmetric Gauss-Laguerre  laser modes and optical lattices may be used to generate toroidal Bose-Hubbard systems \mycite{Amico05} where adjustable weak links may be introduced \mycite{Amico14} to separate the ring into two weakly-coupled subsystems. In this context, one simple configuration may be attained by tilting the lattice potential with respect to a four-node Gauss-Laguerre  mode, thus generating two adjacent high barriers and two adjacent low barriers along the four-site ring, separating it into a trimer and a monomer. Equilibration can be readily detected by monitoring the populations of the two subsystems as a function of time and full relative-number distributions may be attained by multi-realization measurements. As long as the constituent subsystems are weakly-connected, our observations should be independent of the details of the coupling (e.g. which sites of the two subsystems are linked) due to the generic nature of chaotic motion. The interplay between realistic dephasing and particle loss, and the chaotic dynamics will be the subject of future studies.

\section{Discussion}

All stochastic descriptions eventually fail to describe quantum coherent processes, because they inevitably lead to a a microcanonical distribution at $t\rightarrow\infty$, whereas the quantum evolution has an infinite memory of the initial conditions. However, the equivalence between the diagonal and the microcanonical ensembles \mycite{Rigol10,Short12,Reimann12} in the Eigenstate Thermalization Hypothesis (ETH) picture \mycite{Deutsch91, Srednicki94,Rigol08} implies that in the quantum evolution of classically chaotic systems, the memory of initial conditions is effectively lost over an ergodization period with all initial conditions leading to a microcanonical distribution. On longer timescales, quantum recurrences take place and the memory of initial conditions is regained. It is thus understood that stochastic methods should be evaluated by their ability to describe quantum dynamics {\em within  the time scale of interest}, i.e. until an ergodic-like distribution for the pertinent observable is attained.

Within this ergodization time, deviations from LRT include both {\em quantum anomalies}  and {\em semiclassical integrability effects}. The former are directly related to quantization and are important for a dynamical view of Quantum Thermodynamics \mycite{Kozlov}, whereas the latter are related to incomplete chaoticity and residual quasi-integrability regions in the classical mixed phase-space.
 
Our main objective was to highlight a novel quantum anomaly in the 
thermalization process of a quantized chaotic system: a bi-partite Bose-Hubbard complex   
that can be regarded as the building block for thermalization of larger arrays. 
We have demonstrated that thermalization with finite $\hbar$ is quite different 
from that of the corresponding `$\hbar=0$' classical system. 
Whereas classical thermalization is captured well by LRT, leading to an FPE with a Kubo estimate for the energy diffusion coefficient, this approximation fails badly upon quantization. The reason for this dynamical anomaly is the sparsity of the network of couplings between the energy eigenstates of the constituent subsystems which leads to percolation-like dynamics of the energy distribution. As a result, while an FPE description still holds (within the timescale of interest), quantum thermalization, properly described by a resistor-network calculation, can be strikingly slower than the corresponding classical process. \\

\sect{Acknowledgements}
This research has been supported by  by the Israel Science Foundation (grant Nos. 346/11 and 29/11) 
and by the United States-Israel Binational Science Foundation (BSF).

\appendix

\section{Symmetry subspaces in the tetramer} 

The full dimension of the Hilbert space in a tetramer with population $N$ is $\mathcal{N}=(N+1)(N+2)(N+3)/6$. The Hamiltonian of the system can be separated into blocks of smaller dimensions by considering the permutation symmetry between the external trimer sites (sites $2$ and $3$, 
in the schematic illustration inset of \Fig{fig:energyshell}). Denoting the population basis by $\ket{n_0}\ket{n_1,n_2,n_3}$, 
the totally symmetric and the totally anti-symmetric sub-spaces are spanned by the following symmetrized and antisymmetrized superpositions: 
\beq
& \frac{1}{\sqrt{2}} \Big(\ket{n_0}\ket{n_1,n_2,n_3}\pm \ket{n_0}\ket{n_1,n_3,n_2}\Big) \\
& \ket{n_0}\ket{n_1,n,n}.
\eeq
The former is for $n_2\neq n_3$.
We restrict the simulations to the antisymmetric subspace 
which includes less states and therefore allows us to use 
a higher number of particles. 
The antisymmetric subspace excluded the possibility 
of having zero trimer population $x=0$.

\section{Identification of chaos by level statistics}

Given the parameters $N$,$\Omega$,$\omega$,$U$, we find the eigen-energies 
of the Hamiltonian \Eq{eq:hamiltonian} (e.g., \Fig{fig:energyshell}). Dividing the spectrum to small energy intervals, 
we calculate the mean level spacing and the distribution $P(S)$ of level-spacings in each of them. 
We then fit it to the Brody distribution \mycite{Brody81} 
\beq
P_q(S) \ \ = \ \ \alpha S^q\exp(-\beta S^{1+q})
\eeq
with $\alpha=(1+q)\beta$, and $\beta=\Gamma^{1+q}\left[(2+q)/1+q)\right]$. 
Here $\Gamma$ denotes the Euler gamma function. 
A Brody parameter value of $q=0$ indicates a Poissonian level-spacing 
distribution characteristic of the uncorrelated levels of integrable system. 
By contrast for $q=1$ we have the Wigner level-spacing distribution, 
that reflects the level repulsion in the case of a quantized chaotic system. 
Thus, by plotting $q$ as a function of energy
we map the domain of chaotic motion, marked in black in \Fig{fig:energyshell}. The result was then ascertained by inspecting classical Poincare sections in the various regions of the map.

In order to illustrate the connection between the deviation from ergodicity of the saturation profiles and the quasi-integrability islands in the mixed phase-space, we employ the initial states marked in \Fig{fig:energyshell}. Some lie well within the chaotic sea, while others reside in an integrable island. The saturation profiles for these states are shown in \Fig{fig:saturations}, showing a clear connection between integrability and localization.

\section{The resistor-network calculation}

In order to find the diffusion coefficient $D$ for a sparse resistor network 
we rewrite Kirchhoff's law \Eq{eq:kirchhoff} in a matrix form,
\beq
\bm{G} \vec{V}=\vec{I},
\eeq
where $\bm{G}$ is the discrete Laplacian matrix of the network, 
whose diagonal elements are defined as follows: 
\beq
G_{m,m} \ \ \equiv \ \ -\sum_{n'} G_{n',m}.
\eeq
In order to find the conductance of a segment ${[x_1,x_2]}$ 
of length ${dx=x_2{-}x_1}$ 
we shortcut the bonds to the left of the segments, 
hence defining a left lead. Likewise we define a right lead.
Then we place a source $I_{1}=1$ and a sink  $I_{2}=-1$ 
at two nodes on the left and right leads, and solve Kirchhoff's
equation using a psaudo-inverse routine.

Analytical approximation for~$D$ could be obtained 
if the network had well-defined statistical properties.
As an illustrative example we point out that for the 
common model of hopping in a random site network 
the following estimate has been derived \mycite{pts}:   
\beq
D \ \ \approx \ \  
\mathrm{EXP}_{d{+}2}\left(\frac{1}{s}\right)  \  \eexp{-1/s}  \ D_{\tbox{linear}}
\eeq
The polynomial $\mathrm{EXP}_{\nu}(x)$ has degree $\nu$, 
and equals the truncated Taylor expansion of $\exp(x)$.  
Its degree is determined by the effective dimensionality~$d$ of the network.
The linear estimate $D_{\tbox{linear}}$ is what we call here~$D_{\tbox{cl}}$, 
and~$s$ is the sparsity parameter ($s\ll 1$ means sparse network).
If the network originates from the quantization of a weakly 
chaotic system we expect $s$ to be proportional to some power of $1/\hbar$ \mycite{wqc}.
Accordingly the ratio $g_s \equiv D_{\tbox{qm}} / D_{\tbox{cl}}$ 
reflects the sparsity of the network. In the ``sparse" limit (${s\ll 1}$) 
the expression above resembles that of variable-range-hopping.
For small $\hbar$ the network becomes more connected (less sparse) 
and $g_s$ goes to unity. This crossover can be regarded 
as a smoothed ``percolation" transition.  

Form the above discussion it should be clear that sparsity 
and hence the quantum anomaly diminish in the large $N$ limit.  
However, it is important to realize that for thermalization of large arrays,  
which proceeds via progressive process that involves ``chaotic spots" \mycite{Basko}, 
the relevant $\hbar$ is determined by the number of particles per ``spot", 
and not by the total number of particles in the system.

\section{FGR and FPE simulations}

The master equation \Eq{eq:FGR} can be written in a matrix form 
as $(d/dt)\vec{p}=\bm{W}\vec{p}$ and has the solution 
\beq\label{eq:Pt}
\vec{p}(t) \ \ = \ \ \eexp{\bm{W}t} \ \vec{p}(0)~.
\eeq
In order to perform a simulation with the FPE \Eq{eq:FPEx} 
we have to discritize the continuous $x$ variable. 
There are two possible strategies.
One possibility is to define formally a variable~$n$, 
such that $dn/dx=\tilde{g}(x)$. In this variable 
the FPE becomes an unbiased diffusion equation:    
\beq\label{eq:FPEn}
\ppt P_n \ = \ \frac{\partial}{\partial x}\left[D_n \frac{\partial}{\partial x} \Big(P_n \Big) \right],
\eeq
where
\beq\label{eq:Dn}
D_n \ \ = \ \ \tilde{g}(x)^2 D(x).
\eeq
The discrete version of \Eq{eq:FPEn} is a master 
equation with near-neighbor hopping. The rates $D_n$ 
are the same in both directions, and the solution
is straightforward.  
   
The second strategy to solve the FPE, which looks more 
natural in the present context, is to stay with the $x$~variable. 
One should realize that in this variable the ergodic state 
is not uniform. At steady state the current across each $x$ bond
is zero, satisfying
\beq
 w_{x-1,x} \, P(x) \ \ = \ \ w_{x,x-1} \, P(x-1)
\eeq
where $w_{x,x'}$ are transition rates between nodes. 
Selection rules forbid transitions between non-neighboring nodes, 
thus the $\bm{W}$ matrix contains only two diagonals at $x'=x\pm1$.
But unlike the master equation of \Eq{eq:FGR}, 
here $\bm{W}$ is a {\em non-symmetric} matrix. The FPE can thus be viewed as 
a Pauli master equation for $x$ \mycite{Bartsch08}.
At steady state the probability distribution is identical 
to the normalized density of states, hence 
we deduce the relation
\beq
\frac{w_{x-1,x}}{w_{x,x-1}} \ \ = \ \ 
\frac{g(x)}{g(x-1)} \ \ \equiv \ \ \eexp{S}.
\eeq
Accordingly the forward and backward transition rates 
that we are using in the FPE simulation are 
\beq
w_{x-1,x} &=& \left[\frac{S}{1-\exp(-S)}\right] D(x)~,\\
w_{x,x-1} &=& \left[\frac{S}{\exp(S)-1}\right] D(x)~.
\eeq
Using the above rates we can solve the FPE using \Eq{eq:Pt}.

\clearpage
\noindent
{\bf References.-- }

\clearpage

\begin{thebibliography}{99}

\bibitem{Dorfman} 
J.R. Dorfman,
{\em An Introduction to Chaos in Nonequilibrium Statistical Mechanics}. 
Cambridge University Press, Cambridge (1999).

\bibitem{Kinoshita06}
Toshiya Kinoshita, Trevor Wenger, and David S. Weiss,
Nature {\bf 440}, 900 (2006).

\bibitem{Trotzky12}
S. Trotzky,	 Y-A. Chen, A. Flesch, I. P. McCulloch, U. Schollw\"ock, J. Eisert, and I. Bloch
Nat. Phys. {\bf 8}, 325 (2012).


\bibitem{Gring12}
M. Gring, M. Kuhnert, T. Langen, T. Kitagawa, B. Rauer, M. Schreitl, I. Mazets, D. Adu Smith, E. Demler, and J. Schmiedmayer,
Science {\bf 337}, 1318 (2012).

\bibitem{Deutsch91}
J. M. Deutsch, 
Phys. Rev. A {\bf 43}, 2046 (1991).


\bibitem{Srednicki94}
M. Srednicki, 
Phys. Rev. E {\bf 50}, 888 (1994).

\bibitem{Yukalov11}
V. I. Yukalov,
Laser Phys. Lett. {\bf 8}, 485 (2011).

\bibitem{Polkovnikov11}
A. Polkovnikov, K. Sengupta, A. Silva, and M. Vengalattore,
Rev. Mod. Phys. {\bf 83}, 863 (2011).

\bibitem{Rigol08}
Marcos Rigol, Vanja Dunjko, and Maxim Olshanii,
Nature {\bf 452}, 854 (2008)  

\bibitem{Cassidy09}
Amy C. Cassidy, Douglas Mason, Vanja Dunjko, and Maxim Olshanii,
Phys. Rev. Lett. {\bf 102}, 025302 (2009)  

\bibitem{Eckstein08}
M. Eckstein and M. Kollar,
Phys. Rev. Lett. {\bf 100}, 120404 (2008).


\bibitem{Pal10}
A. Pal and D. A. Huse,
Phys. Rev. B {\bf 82}, 174411 (2010).

\bibitem{Ponomarev11}
A. V. Ponomarev, S. Denisov, and P. H\"anggi,
Phys. Rev. Lett. {\bf 106}, 010405 (2011).



\bibitem{Bunin11}
Bunin, G., D'Alessio,  L.,  Kafri, Y. \& Polkovnikov, A.
Nature Physics {\bf 7}, 913 (2011).

\bibitem{Ates12}
C. Ates, J. P. Garrahan, and I. Lesanovsky,
Phys. Rev. Lett. {\bf 108}, 110603 (2012).

\bibitem{Tikhonenkov13}
I. Tikhonenkov, A. Vardi, J. R. Anglin, and D. Cohen,
Phys. Rev. Lett. {\bf 110}, 050401 (2013).

\bibitem{Niemayer13}
H. Niemeyer, D. Schmidtke, and J. Gemmer,
Euro. Phys. Lett. {\bf 101}, 10010 (2013).

\bibitem{Niemayer14}
H. Niemeyer, K. Michielsen, H. De Raedt, and J. Gemmer,
Phys. Rev. E {\bf 89}, 012131 (2014).

\bibitem{Bartsch08}
C. Bartsch, R. Steinigeweg, J. Gemmer,
Phys. Rev. E {\bf 77}, 011119 (2008)




\bibitem{Fleishman80}
L. Fleishman and P. W. Anderson,
Phys. Rev. B {\bf 21}, 2366 (1980).

\bibitem{Basko06}
D. Basko, I. L. Aleiner, and B. Altshuler,
Ann. Phys.  {\bf 321}, 1126 (2006).

\bibitem{BarLev14}
Y. Bar Lev and D. R. Reichman, 
Phys. Rev. B 89, 220201 (2014).




\bibitem{Wilk88}
M. Wilkinson, 
{J. Phys. A} {\bf 21}, 4021 (1988).

\bibitem{Wilk95}
M. Wilkinson, E.J. Austin,
{J. Phys. A} {\bf 28}, 2277 (1995).

\bibitem{Cohen99}
D. Cohen, 
Phys. Rev. Lett. {\bf 82}, 4951 (1999).




\bibitem{Yurovsky11}
V. A. Yurovsky and M. Olshanii,
PRL {\bf 106}, 025303 (2011).

\bibitem{Santos12}
L. F. Santos, F. Borgonovi, and F. M. Izrailev,
PRL {\bf 108}, 094102 (2012)

\bibitem{slk}
D. Cohen, 
Physica Scripta T151, 014035 (2012). 

\bibitem{Chuchem10}
M. Chuchem, K. Smith-Mannschott, M. Hiller, T. Kottos, A. Vardi, and D. Cohen,
Phys. Rev. A {\bf 82}, 053617(2010).

\bibitem{Kottos} 
M. Hiller, T. Kottos, and T. Geisel, 
Phys. Rev. A {\bf 79}, 023621 (2009); 
{\em and further references therein}. 

\bibitem{Basko} 
D.M. Basko, Ann. Phys. 326, 1577 (2011)

\bibitem{Henn}
H. Hennig, R. Fleischmann, 
Phys. Rev. A 87, 033605 (2013)

\bibitem{Brody81}
T.A. Brody, J. Flores, J.B. Fench, P.A. Mello, A. Pandey, and S.S.M. Wong, 
Rev. Mod. Phys. {\bf 53}, 385 (1981).




\bibitem{Rigol10}
M. Rigol, L.F. Santos,
Phys. Rev. A {\bf 82}, 011604(R) (2010).

\bibitem{Short12}
A.J. Short, T.C. Farrelly,
New J. Phys. {\bf 14}, 013063 (2012).

\bibitem{Reimann12}
P. Reimann, M, Kastner,
New J. Phys. {\bf 14}, 043020 (2012).

\bibitem{Amico05}  
L. Amico, A. Osterloh, and F. Cataliotti,
Phys. Rev. Lett. {\bf 95}, 063201 (2005).

\bibitem{Henderson09}
K. Henderson, C. Ryu, C. MacCormick, M.G. Boshier,
New J. Phys. {\bf 11} 043030 (2009).

\bibitem{Wright13}
K.C. Wright, R.B. Blakestad, C.J. Lobb, W.D. Phillips, G.K. Campbell, 
Phys. Rev. Lett. {\bf 110}, 025302 (2013).

\bibitem{Amico14}
L. Amico,D. Aghamalyan, F. Auksztol,H. Crepaz,R. Dumke, L.C. Kwek
Sci. Rep. {\bf 4}, 4298 (2014).

\bibitem{Kozlov}
R. Kosloff,
{\em Entropy} {\bf 15}, 2100 (2013).



\bibitem{pts}
Y. de Leeuw, D. Cohen,
Phys. Rev. E 86, 051120 (2012)

\bibitem{wqc}
A. Stotland, L.M. Pecora and D. Cohen, 
Phys. Rev. E 83, 066216 (2011)

\end{thebibliography}
\end{document}